\documentstyle{article}

\catcode`\@=11

\def\@citex[#1]#2{\if@filesw\immediate\write\@auxout
	{\string\citation{#2}}\fi
\def\@citea{}\@cite{\@for\@citeb:=#2\do
	{\@citea\def\@citea{,}\@ifundefined
	{b@\@citeb}{{\bf ?}\@warning
	{Citation `\@citeb' on page \thepage \space undefined}}
	{\csname b@\@citeb\endcsname}}}{#1}}
\def\@cite#1#2{{$\null^{#1}$\if@tempswa\typeout
	{warning: optional citation argument 
	ignored: `#2'} \fi}}

\renewenvironment{thebibliography}[1]
	{
	 \begin{list}{${}^{\arabic{enumi}}\!\!$}
	{\usecounter{enumi}
\setlength{\leftmargin 7pt}{\rightmargin 0pt}   
\sloppy}}
{\end{list}}

\newcommand{\nonumsection}[1] {\vspace{12pt}\noindent{\large \bf #1}
	\par\vspace{5pt}}

\newtheorem{lem}{Lemma}
\newtheorem{prop}{Proposition}
\newtheorem{cor}{Corollary}
\newtheorem{theo}{Theorem}

\def\ba{\begin{eqnarray}}
\def\ea{\end{eqnarray}}
\def\be{\begin{equation}}
\def\ee{\end{equation}}

\def\Z{\hbox{\msbm Z}}

\def\R{\hbox{\msbm R}}
\def\N{\hbox{\msbm N}}
\def\Ni{\hbox{\msbms N}}
\def\C{\hbox{\msbm C}}
\def\Q{\hbox{\msbm Q}}

\def\ag{{{\cal A}/{\cal G}}}
\def\agb{{\overline \ag}}
\def\mal{\mu_{AL}}
\def\a{\alpha}

\def\s{\sigma}
\def\S{\Sigma}
\def\wt{\widetilde}

\def\hg{{\cal HG}}
\def\Ab{\mbox{$\bar A$}}

\def\ms{\mbox{$\mu_{\rho}$}}

\def\cs{{\cal S}}
\def\vff{\varphi}
\def\ss{\cs(\R^{d+1})}
\def\csa{\cs^a(\R^{d+1})}

\def\l2m{L^2(M,\mu)}
\def\l1m{L^1(M,\mu)}
\def\f{\mbox{$\phi$}}
\def\ns{\mbox{$\mu_{\sigma}$}}
\def\sd{\mbox{$\cs ({\R}^{d+1})$}}
\def\sdd{\mbox{$\cs^{'} ({\R}^{d+1})$}}

\def\nm{\mbox{$\mu_{C_m}$}}
\def\vf{\mbox{$\varphi$}}

\newfont{\msbm}{msbm10}
\newfont{\msbms}{msbm6}  

\begin{document}


\title{
Physical Properties of \\ Quantum Field Theory Measures}

\author{J.M. Mour\~ao${}^{\rm a)}$, T. Thiemann${}^{\rm b)}$ 
and J.M. Velhinho${}^{\rm c)}$}

\date{}

\maketitle

\begin{abstract}
Well known methods of measure theory on infinite dimensional
spaces are used to study physical properties of measures
relevant to quantum field theory. The difference of typical
configurations of free massive scalar field theories with
different masses is studied.
We apply the same methods to study the Ashtekar-Lewandowski (AL)
measure on spaces of connections. In particular we
prove that the diffeomorphism group acts
ergodically, with respect to the
AL measure, on the Ashtekar-Isham space of quantum
connections modulo gauge transformations. We also 
prove that a typical, with respect to the AL measure,
quantum connection restricted to a (piecewise analytic)
curve leads to a parallel transport discontinuous at every
point of the curve.

   
\vskip .5 in
\centerline{\large Published in: J. Math. Phys. 40 (1999) 2337} 

\end{abstract}


\newpage 

\pagestyle{myheadings}
\markboth{Quantum Field Theory Measures}{Quantum Field Theory Measures}



\section{Introduction}
\label{s1}

Path integrals play an important role in modern
quantum field theory. The application in this context of methods
of the mathematical theory of measures on infinite dimensional
spaces is due to constructive quantum field
theorists\cite{sim}${}^{\hbox{-}}$\cite{AJPS}.
With the help of these methods 
important physical results have been obtained, especially
concerning two and three dimensional theories.
Recently analogous methods have been applied
within the framework of Ashtekar non-perturbative
quantum gravity to give a rigorous meaning
to the connection representation\cite{ashish}${}^{\hbox{-}}$\cite{lew},
solve the diffeomorphism constraint\cite{almmt} and
define the Hamiltonian constraint\cite{tho1}${}^{\hbox{-}}$\cite{tho4}.
These works all crucially depend on the use of (generalized) Wilson loop 
variables
which have been considered for the first time, within a Hamiltonian 
formulation of gauge theories, by Gambini, Trias and 
collaborators\cite{Gam1,Gam2,Gam3}
and were rediscovered for canonical quantum gravity by 
Rovelli and Smolin\cite{l1}. 
In fact, instead of working in the connection representation 
for which Wilson loops are just convenient functions, one can use a so-called
loop representation (see Ref.~23
and references therein) by means of which 
a rich arsenal of (formal) results was obtained which complement those 
obtained in the connection representation. For
previous works on measures on spaces of connections see e.g. Refs.~24 and 25.

The present paper has two main goals.
The first consists
in studying
physical properties of the support of the path integral
measure of free massive scalar fields. We
use a system with a countable number of simple random variables
which probe the typical scalar fields over cubes,
with volume $L^{d+1}$, placed far away
in (Euclidean) space-time. With these probes we are able
to study the difference between the supports of two
free scalar field theories with different masses.
The results that we obtain provide a characterization of the
supports which is physically more transparent than those
obtained previously\cite{collan}${}^{\hbox{-}}$\cite{Ku}.

Our second goal consists in providing the proof of analogous
results for the Ashtekar-Lewandowski (AL) measure on the
space $\agb$ 
of quantum connections modulo gauge transformations\cite{ashish,ashlew}. 
First we prove that the group of diffeomorphisms acts
ergodically, with respect to the AL measure, on $\agb$. Second
we show that the AL measure is supported on connections which,
restricted to a curve, lead to parallel transports discontinuous at every
 point of the curve.

Quantum scalar field theories have been intensively studied
both from the mathematical and the physical point of view.
Divergences in Schwinger functions (which, in measure theoretical
terminology, are the moments of the measure) are directly related
with the fact that the relevant measures are supported 
not on the space of nice smooth scalar field configurations,
that enter the classical action, but rather on spaces of distributions.
This gives a strong motivation for more detailed studies
of the support of relevant measures. In physical terms this
corresponds to finding ``typical (quantum) scalar field configurations'',
the set of which has measure one.

The present paper is organized as follows. In section \ref{s2} we
recall some results from the theory of measures on infinite 
dimensional spaces. Namely Bochner-Minlos theorems are 
stated and the concept of ergodicity of (semi-)group actions
is introduced. In section \ref{novos3}
we study properties of simple measures: 
the countable product of identical one-dimensional Gaussian measures 
and the white noise measure. In the first example we illustrate
the disjointness of the support of different measures, which are 
invariant under a fixed ergodic action of the same group. 
For the white noise measure we choose random variables which probe 
the support and which will be also used in section \ref{s3}
to study the support of massive scalar field theories.
In section \ref{s4} we obtain
properties of  the
support of the AL measure that complement previously obtained 
results\cite{marmou} and prove that the diffeomorphism group
acts ergodically on the space of connections modulo gauge 
transformations.  
In section \ref{s6} we present our conclusions.


\section{Review of Results From Measure Theory}
\label{s2}


\subsection{Bochner-Minlos Theorems}
\label{s2.1}

In the characterization of typical configurations
of measures on functional spaces the so called
Bochner-Minlos theorems play a very important
role. These theorems are  infinite dimensional generalizations of the
Bochner theorem for probability measures on $\R^N$.
Let us, for the convenience of the reader, recall the
latter result. Consider any (Borel) probability measure
$\mu$ on $\R^N$, i.e. a finite measure, normalized
so that $\mu(\R^N)=1$. The generating functional $\chi_\mu$ of this measure
is its Fourier transform, given by the following  function on $\R^N$
($\cong (\R^N)'$, the prime denotes the topological dual, see below) 
\be
\label{2.1}
\chi_\mu(\lambda)= \int_{\R^N} d\mu(x) e^{i(\lambda ,  x)} \ ,
\ee  
where $(\lambda ,  x) = \sum_{j=1}^N \lambda^j x_j$.
Generating functionals of measures satisfy the 
following three basic conditions,
\begin{itemize}

\item[({\it i}\/)] Normalization: $\chi(0) = 1$;

\item[({\it ii}\/)] Continuity: $\chi$ is continuous on $\R^N$;

\item[({\it iii}\/)] Positivity: $\sum_{k,l=1}^m c_k \overline{c_l} 
\chi(\lambda_k - \lambda_l) \geq 0$, for all $m \in \N$, $c_1, ... ,
c_m \in \C$ and $\lambda_1 , ... , \lambda_m \in \R^N$.

\end{itemize} 
The last condition comes from the fact that $||f||_\mu \geq 0$,
for $f(x) = \sum_{k}^m c_k e^{i(\lambda_k,x)}$,
where $|| \cdot ||_\mu$ denotes the $L^2(\R^N,d\mu)$ norm. 
The finite dimensional Bochner theorem states that
the converse is also true. Namely, for any function
$\chi$ on $\R^N$ satisfying ({\it i}\/), ({\it ii}\/) and ({\it iii}\/)
there exists a unique probability measure on $\R^N$ 
such that $\chi$ is its generating functional.

Both in statistical mechanics and in
quantum field theory one is interested
in the so-called correlators, or in 
probabilistic terminology, the moments
of the measure $\mu$,
\be
\label{2.2}
<(x_{i_1})^{p_1} ... \ (x_{i_k})^{p_k}> := \int_{\R^N}d\mu(x)
(x_{i_1})^{p_1} ... \  (x_{i_k})^{p_k} \ .
\ee 
For the correlators of any order to exist
the measure $\mu$ must have a rapid decay at $x-$infinity,
in order to compensate the polynomial growth in (\ref{2.2})
(examples are Gaussian
measures and measures with compact support). In
the $\lambda$-space the latter condition turns out to
be equivalent to $\chi$ being infinitely differentiable ($C^\infty$).
The correlators are then just equal to partial derivatives
of $\chi$ at the origin, multiplied by an appropriate
power of $-i$.

Let us now turn to the infinite dimensional case.
The role of the space
of $\lambda$'s will be played by ${\cal S} (\R^{d+1})$,
the Schwarz space of  $C^\infty$-functions on
(Euclideanized) space-time with fast decay at infinity.
So we have
the indices $\lambda(i) := \lambda^i \ $  replaced by $ \ f(x)$.
The space  ${\cal S} (\R^{d+1})$ has a standard (nuclear) topology.
Its elements are functions with regularity 
properties both for  small and for large distances.
The physically interesting measures will
``live'' on spaces dual to  ${\cal S} (\R^{d+1})$. 
Consider the space $ {\cal S}^{'} (\R^{d+1})$ of all
continuous linear functionals on ${\cal S} (\R^{d+1})$
(i.e. the topological dual of  ${\cal S} (\R^{d+1})$).
This is the so called space of tempered
distributions, which includes delta functions
and their derivatives, as well
as functions which grow polynomially
at infinity.
We will consider also the even bigger space
${\cal S}^a(\R^{d+1})$
of all linear (not necessarily continuous)
functionals on ${\cal S}(\R^{d+1})$. 
Then the simplest generalization of
the Bochner theorem states that
a function $\chi(f)$ on $\ss$ 
satisfies the following conditions,
\begin{itemize}

\item[({\it i'}\/)] Normalization: $\chi(0) = 1$;

\item[({\it ii'}\/)] Continuity: $\chi$ is continuous on any
finite dimensional subspace of $\ss$;

\item[({\it iii'}\/)] Positivity: $\sum_{k,l=1}^m c_k \overline{c_l} 
\chi(f_k - f_l) \geq 0$, for all $m \in \N$, $c_1, ... ,
c_m \in \C$ and $f_1 , ... , f_m \in \ss$,

\end{itemize} 
if and only if it is the Fourier transform
of a probability measure $\mu$ on $\csa$,
i.e.
\be
\label{2.3}
\chi(f)= \int_{\csa} d\mu(\phi) e^{i\phi (f)} \ \ .
\ee

The topology of convergence on finite dimensional
subspaces of $\ss$ is unnaturally strong.
Demanding in ({\it ii'}\/) continuity of $\chi$ with respect to the
much weaker standard nuclear topology on $\ss$
yields a measure supported on the topological dual
$\cs^{'} (\R^{d+1})$ of $\ss$\cite{foot1}.
This is the first
version of the Bochner-Minlos theorem. 
Further refinement can be achieved 
if $\chi$ is continuous with
respect to a even weaker topology induced by an inner product.
We present a special version of this result, suitable for the 
purposes of the present work; for different, more general 
formulations see Refs.~3 and 32.
Let $P$ be a linear continuous
operator from $\ss$ onto $\ss$, with continuous inverse.
Suppose further that $P$ is positive when viewed as an operator
on $L^2({\R}^{d+1},d^{d+1}x)$ and that the bilinear form 
\be
\label{2.3b}
< f_1 , f_2 >_{P^{1/2}}\, := (P^{1/2} f_1, P^{1/2} f_2)\ ,\ \ \
f_1, f_2 \in \ss
\ee
defines an inner product on $\ss$, where $( \ , \ )$ denotes
the $L^2({\R}^{d+1},d^{d+1}x)$ inner product.
Let $\chi$ satisfy ({\it i'}\/), ({\it iii'}\/) and be continuous 
with respect to
the norm associated with the inner product
$< \ , \ >_{P^{1/2}}$. Natural examples are provided by
Gaussian measures $\mu_C$ with covariance $C$
and Fourier transform 
\be
\label{2.3c}
\chi_C(f) = e^{-{1 \over  2} (f, C f)}
\ ,
\ee
in which case one can take the positive operator $P$ to be the 
covariance $C$ itself.
A particular case is the path integral measure for free massive scalar
fields with mass $m$, which is the Gaussian measure
with covariance
\be
\label{2.3d}
C_m = (-\Delta + m^2)^{-1} 
\ ,
\ee
where $\Delta$ denotes the Laplacian on $\R^{d+1}$.
In the general (not necessarily Gaussian) case let
${\cal H}_{P^{1/2}}$ 
(${\cal H}_{P^{-1/2}}$)
denote the completion of $\ss$ with respect to the inner product
$< \ , \ >_{P^{1/2}}$ ($< \ , \ >_{P^{-1/2}}$).
Then the measure on $\cs^{'}(\R^{d+1})$ corresponding to
$\chi$ is actually supported on a proper
subset of  $\cs^{'}(\R^{d+1})$ given by an
extension of ${\cal H}_{P^{-1/2}}$ defined by a Hilbert-Schmidt
operator on ${\cal H}_{P^{1/2}}$.
We see that in the scalar field case  ${\cal H}_{C_m^{-1/2}}$
is the space of finite action configurations and therefore
typical quantum configurations live in a bigger space.
In order to define the extension mentioned above,
recall that an operator $H$ on 
a Hilbert space is said to be Hilbert-Schmidt
if given an (arbitrary) orthonormal basis $\{e_k\}$ one has
$$
\sum_{k=1}^{\infty} <\!He_k,He_k\!> \ \ \ <\  \infty   \ .
$$
Given such a Hilbert-Schmidt operator $H$ on ${\cal H}_{P^{1/2}}$,
which we require to be invertible, self-adjoint
and such that $H(\ss)\subset\ss$, define the new inner product
$< \ , \ >_{P^{-1/2}H}$ on $\ss$ by
\be
\label{supp}
< f_1 , f_2 >_{P^{-1/2}H}\, := (P^{-1/2} H f_1, P^{-1/2} H f_2)\ ,\ \ \
f_1, f_2 \in \ss .
\ee
Consider ${\cal H}_{P^{-1/2}H}$, the completion of $\ss$ with respect to the
inner product $< \ , \ >_{P^{-1/2}H}$, and identify its elements with
linear functionals on $\ss$ through the $L^2({\R}^{d+1},d^{d+1}x)$
inner product. Under the above conditions, the (second version of the)
Bochner-Minlos Theorem
states that
\medskip

\begin{itemize}

\item[] {
\bf  (Bochner-Minlos)} 
A generating functional $\chi$, continuous with respect to
the inner product $<\ , \ >_{P^{1/2}}$, 
is the Fourier transform of
a unique measure supported on 
${\cal H}_{P^{-1/2}H}$, for every Hilbert-Schimdt operator  $H$ on
${\cal H}_{P^{1/2}}$ such that  $\ss\subset {\rm Ran}H,\ H^{-1}(\ss)$ 
is dense 
and $H^{-1}:\ss\rightarrow {\cal H}_{<\ , \ >_{P^{1/2}}}$
is continuous.
\end{itemize}

\noindent In section \ref{s2.3} we will also use an obvious adaptation
of this result to the space of infinite
sequences $\R^{\Ni}$.

A common feature of the two versions of the Bochner-Minlos
theorem is that they give the support as a linear
subspace of the original measure space.
Nonlinear properties of the support have
to be obtained in a different way. In particular
one can show (see section \ref{s2.4})
that the white noise measures with
\ba
\label{2.4}
\chi_{\s_1}(f)\!\! & := & \!\! e^{-{\s_1 \over 2}(f,f)}   \\
\noalign{\hbox{\rm and}}
 \chi_{\s_2}(f)\!\!  & := &\!\! e^{-{\s_2 \over 2}(f,f) }   
\ea
have disjoint supports for $\s_1, \s_2 > 0$ and  $\s_1 \neq \s_2$,
while
Bochner-Minlos theorem would give the same results
in both cases.


\subsection{Ergodic Actions}
\label{s2.2}

We review here some concepts and results from ergodic
theory\cite{yam,sin}. Let $\varphi$ denote an action
of the group $G$
on the space $M$, endowed with a
probability measure $\mu$, by measure preserving
transformations $\varphi_g : M \rightarrow M \ , \  g \in G$,
i.e. $\varphi_{g_*} \mu = \mu \ ,  \quad \forall g \in
G$ or, equivalently, for every measurable set   $A \subset M$
and for every $g \in G$, 
the measure of $A$ equals the measure of the pre-image of $A$ by $\varphi_g$.
The action $\varphi$ is said to be ergodic
if all $G$-invariant sets have either
measure zero or one. The fact that $\varphi$
is measure preserving implies that the (right)
linear representation $U$ of $G$ on
$L^2(M,d\mu)$ induced by $\varphi$
\be
\label{2.5}
(U_g\psi)(x) := \psi(\varphi_{g}x)      \ 
\ee
is unitary.
The action $\varphi$ on $M$ is ergodic
if and only if the only $U_G$-invariant vectors
on $L^2(M,d\mu)$ 
are the (almost everywhere) constant functions. This follows
easily from the fact that the
linear space spanned by characteristic
functions of measurable sets (equal to one on the set
and zero outside) is dense in $L^2(M,d\mu)$.
The above also applies for a discrete semi-group generated by a single
(not necessarily invertible) measure preserving transformation $T$, 
in which case the role of the group $G$ is played by the additive 
semi-group \N, $T$ being identified with $\varphi_1$. Notice
that in the latter case the linear representation on $L^2(M,d\mu)$ may
fail to be unitary, although isometry still holds.
For  actions of $\R$ and $\N$, respectively,
the following properties are equivalent to ergodicity\cite{sin}
$$
\lim_{\tau \rightarrow \infty}{1 \over 2\tau} \int_{-\tau}^\tau dt
\psi(\varphi_t x_0)  =  \int_M d\mu(x) \psi(x)   \eqno(11{\rm a})
$$
$$
\lim_{N \rightarrow \infty}{1 \over  N+1} \sum_{n=0}^N
\psi(\varphi_n x_0)  =  \int_M d\mu(x) \psi(x)   \ ,  \eqno(11{\rm b})
$$
\addtocounter{equation}{1}
where $\varphi_n := \varphi_1^n$ and
the equalities hold for $\mu$-almost every $x_0$
and for all $\psi \in L^1(M,d\mu)$. One important
consequence of (11) is that if
$\mu_1$ and $\mu_2$ are  two
different measures and a given action
$\varphi$ is ergodic with respect to both $\mu_1$
and $\mu_2$ then these measures must have
disjoint supports (points $x_0$ for
which (11) holds). Recall also that
the action of $\R$ and $\N$, respectively, is called mixing
if for every $\psi_1, \psi_2 \in L^2(M,d\mu)$ we  have
$$
\lim_{t \rightarrow \infty}
<\psi_1, U_t \psi_2> 
\  =  \ <\psi_1, 1>
<1, \psi_2> \, , \eqno(12{\rm a})
$$
$$
\lim_{n \rightarrow \infty}
<\psi_1, U_n \psi_2> 
\ =  \ <\psi_1, 1>
<1, \psi_2>     \, . \eqno(12{\rm b})
$$
\addtocounter{equation}{1}
It follows from (12) that every $U$-invariant  
$L^2(M,d\mu)$-function is constant almost everywhere
and therefore every mixing action is ergodic 
(see Ref.~33
for details). 
If $M$ is a linear space then (11) gives
a non-linear characterization of the
support. Indeed if $x_1$ and $x_2$ are
typical configurations, in the sense
that (11)  holds for them, then
$x_1 + x_2$ and $\lambda x_1$ for
$\lambda \neq 1$, are in general not
typical configurations. The nonlinearity of supports
is best illustrated by the action of
$\N$
on the space of infinite sequences
endowed with a Gaussian measure that we
recall in the next subsection.


\section{Support Properties of Simple Measures}
\label{novos3}


\subsection{Countable Product of Gaussian Measures}
\label{s2.3}

We will consider here the simplest case of a Gaussian
measure in an infinite dimensional space.
We will see however that many aspects of Gaussian
measures on functional spaces can be
rephrased in this simple context.

Let   $M=\R^{\Ni}$, the set of all
real sequences (maps from $\N$ to $\R$)
$$
x=\{x_1,x_2,\ldots\}
$$
and consider on this space the measure given by  
the infinite product of identical Gaussian measures on $\R$, 
of mean zero and variance $\rho$\
\be
\label{3.5}
d\ms(x)=\prod_{n=1}^{\infty}e^{-{x_n^2 \over 2\rho}}
{dx_n \over \sqrt{2\pi\rho}}\ .
\ee
As we saw above, an equivalent way of defining $\ms$ is by giving 
its  Fourier Transform.
Let 
\be
\label{inner}
{<y,z>}_{\sqrt{\rho}} := 
\rho \, (y , z) \ ,\ \
\  y,z \in {\cal S} \ ,
\ee
where 
$(y , z)=\sum_{n = 1}^{\infty} y_n z_n$ and  
$\cal S$ is the space of rapidly decreasing sequences
$$
{\cal S}:=\Bigl\{y\in\R^{\Ni}\ :\quad
\sum^{\infty}_{n=1}n^k y_n^2<\infty\ ,\ 
\forall k>0\Bigr\}\ .$$
Then
\be
\label{3.6}
\chi_{\rho}(y) := e^{-{1 \over 2}{<y,y>}_{\sqrt{\rho}}} =
\int
e^{i(y,x)}d\ms(x) \ ,
\ee
where $y\in {\cal S}$, $x\in \R^{\Ni}$.

Consider now the ergodic (in fact mixing) action $\vff$ of
$\N$ generated by $T\equiv\vff_1$, 
\be
\label{3.7}
(\vff_1(x))_n := x_{n+1} \ , \ \ x \in     \R^{\Ni}
\ee
which, as we will see, is a discrete analogue of the
action of $\R$ by translations on
the quantum configuration space of a free
scalar field theory.
The transformation $\vff_1$
is clearly measurable and measure preserving,
since all the measures in the product (\ref{3.5}) are equal. 
It is also not 
difficult   to see that $\vff$ is mixing. In fact, invoking linearity and
continuity, one only needs to show that (12b) 
is verified for a set of
$L^2$-functions whose span is dense. The functions of
the form $\exp (i(y, x))$, $y \in {\cal S}$ form such a set, 
and one has for them
\begin{eqnarray*}
\lim_{n \to \infty}<e^{i\, (y, x)},U_n 
e^{i\, (z, x)}>\! & = &
\! e^{-{1 \over 2}
\left( {<z,z>}_{{\sqrt{\rho}}}+{<y,y>}_{{\sqrt{\rho}}} \right)} \\
& = & \! <e^{i\, (y, x)},1><1,e^{i\, (z, x)}>\ ,
\ \ \forall y,z \in {\cal S}   \ ,
\end{eqnarray*}
where, in the first and last terms, $< \ , \ > $
denotes the $L^2(\R^{\Ni},d\ms)$ inner product and 
$U$ denotes the isometric representation associated with $\vff$.
Thus, $\vff$ is an ergodic action with respect to\ $\ms$, 
for any $\rho$.
Of course, $\rho \not = \rho'$ implies $\ms \not =
\mu_{\rho'}$ and one concludes
that $\ms$ and $\mu_{\rho'}$ 
must be mutually singular (i.e. have disjoint supports).
In fact, taking in (11b) $\, \psi(x)=\exp (i(y,x))$
for both $\ms$ and $\mu_{\rho'}$ 
leads to a contradiction,
unless a $x_0$ satisfying (11b) 
for both $\ms$ and $\mu_{\rho'}$
cannot be found.

The aim now is to find properties of typical configurations which allow
to distinguish the supports of $\ms$ and 
$\mu_{\rho'}$.
Unfortunatelly the Bochner-Minlos theorem cannot help, 
due to the fact that the inner products (\ref{inner}),
that define the measures 
$\ms$ and $\mu_{\rho'}$ via (\ref{3.6}),
are proportional to each other
and therefore the corresponding extensions in the 
Bochner-Minlos theorem are equal: 
${\cal H}_{\rho^{-1/2}H}={\cal H}_{{\rho'}^{-1/2}H}\ (={\cal H}_H)$
for any $\rho,\rho'$.

Let us now find a better characterization of the support, for which
the mutual singularity of  $\ms$ and $\mu_{\rho'}$
becomes explicit. In order to achieve this we use a slight modification of an 
argument given in Ref.~1,
that provides convenient sets, both
of measure zero and one. 
\begin{prop}
\label{prop1}
Given a  sequence $\{\Delta_j \}$, $\Delta_j>1$ the
$\ms$-measure of the set
\be
\label{3.20}
\quad Z_{\rho}(\{\Delta_j\})
:=\{x\ :\quad \exists N_x \in \N \quad \mbox{\em s.t.}\quad |x_n|<
\sqrt{2\rho\ln \Delta_n}\, ,
\  \mbox{\em for}\  n\geq N_x  \}
\ee
is one (zero) if $\ \sum 1/\bigl(\Delta_j\sqrt{\ln\Delta_j}\bigr)$
converges (diverges).
\end{prop}
This can be proven as follows. 
For fixed integer $N$ and positive sequence $\{\Lambda_j\}$
define sets $Z_N(\{\Lambda_j\})$ by
\be
\label{3.9}
Z_N(\{\Lambda_j\})
:=\{x\ :\quad
|x_n|<\Lambda_n\ ,\  \mbox{for}\  n\geq N  \}
\ee
The $\ms$-measure of each of these sets is
\begin{eqnarray}
\label{3.14}
\ms (Z_N(\{\Lambda_j\})) & = &
\prod^{\infty}_{n=N}\mbox{Erf}\left(
{\Lambda_n \over \sqrt{2\rho}}
\right)    \ ,
\end{eqnarray}
where Erf$(x) = 1/\sqrt{\pi} \int_{-x}^x e^{-\xi^2} d\xi $ is
the error function.
The sequence of sets $Z_N(\{\Lambda_j\})$
is an increasing sequence for fixed $\{\Lambda_j\}$, and
the set $Z_{\rho}(\{ \Delta_j \})$ defined
in (\ref{3.20})  is just their infinite union, for
$\Delta_j=\exp \bigl(\Lambda_j^2 / 2\rho\bigr)\, $:
\be
\label{3.18}
Z_{\rho}(\{\Delta_j\}) = \bigcup_{N\in\Ni}             
      Z_N(\{\Lambda_j\})    \ .                   
\ee
$ $From $\s$-additivity  one gets            
\be
\label{3.13}
\ms (Z_{\rho}(\{\Delta_j\}))               
=        
\lim_{N\to \infty}\,
\ms (Z_N(\{\Lambda_j\}))             
\ee
and therefore
\be
\label{3.15}
\ms (Z_{\rho}(\{\Delta_j\}))=\exp\left(\lim_{N \to \infty}
\sum^{\infty}_{n=N}\ln(\mbox{Erf}(\sqrt{\ln\Delta_n}))\right) \ .
\ee
Notice that the exponent is the limit of the
remainder of order $N$ of a series. Since
only divergent sequences $\{ \Delta_j\}$
may lead to a non-zero measure, one can use
the asymptotic expression for Erf$(x)$,
which gives
\be
\label{3.15b}
\ms (Z_{\rho}(\{\Delta_j\}))=\exp\left( -
\lim_{N \to \infty}
\sum^{\infty}_{n=N}{1 \over \Delta_n \sqrt{\ln\Delta_n}}\right)\ .
\ee
Depending on the sequence $\{\Delta_j\}$,              
only two cases are
possible: either the series exists, or it diverges to plus infinity,
since $\Delta_j>1\, , \forall j$. In the first case the limit 
of the remainder is
zero, and so the measure of
$Z_{\rho}(\{\Delta_j \})$ will be one. If the sum diverges
so does the 
remainder of any order and therefore the measure of
$Z_{\rho}( \{ \Delta_j \})$ is zero.~$\Box$

Let us now discuss the meaning of this result.
To begin with, it is easy to present disjoint sets $A$, $A'$ s.t 
$\ms(A)=1$, $\mu_{\rho'}(A)=0$ and 
$\ms(A')=0$, $\mu_{\rho'}(A')=1$, for $\rho \not = \rho'$.
Without loss of generality, take $\rho=a \rho' ,\, a>1$.
The set $Z_{\rho}(\{n\})$
has $\ms$-measure zero, since $\sum 1/(n \sqrt{\ln n})$ 
diverges. But $Z_{\rho}(\{n\})=
Z_{\rho'}(\{n^a\})$ (see (\ref{3.20}))
and $Z_{\rho'}(\{n^a\})$ has $\mu_{\rho'}$-measure one,
since $\sum 1/(n^a\sqrt{\ln n^a})$ converges, for $a>1$.
On the other hand  the sets 
$Z_{\rho}(\{n^{1+\epsilon}\})$
are such that
$$
\ms (Z_{\rho}(\{n^{1+\epsilon}\})) = 
\mu_{\rho'}(Z_{\rho}(\{n^{1+\epsilon}\})) = 1 \ ,\ \ \forall\epsilon > 0
$$
and since $Z_{\rho}(\{n\}) \subset Z_{\rho}(\{n^{1+\epsilon}\}),\ 
\forall\epsilon > 0$,
the difference sets
\be
\label{difset}
A_{\rho}^{\epsilon} :=  Z_{\rho}(\{n^{1+\epsilon}\}) \backslash
Z_{\rho}(\{n\}) 
\ee
are such that $\ms (A_{\rho}^{\epsilon}) = 1$ 
and $\mu_{\rho'}(A_{\rho}^{\epsilon}) = 0\, ,\ \forall\epsilon>0$.

Notice that the ``square-root-of-logarithm'' nature of the support 
$A_{\rho}^{\epsilon}$ of $\ms$ 
does not
mean that the typical sequence $x$  approaches
$\sqrt{2\rho\ln n}$, as ${n\to \infty}$. 
To clarify this point let us appeal to the Bochner-Minlos Theorem
(see section \ref{s2.1}). Take ${\cal H}_{P^{1/2}}$ to be
$\ell^2$, the completion of $\cal S$ with respect to the inner product 
$(\ ,\ )\ $:
$$
\ell^2 = \Bigl\{y \in \R^{\Ni} \ :
\ \sum_{n=1}^{\infty}y_n^2  \ < \ \infty \Bigr\} \, .
$$
Consider a vector $a\!=\!\{a_1,a_2,\ldots\}\in\ell^2$
and a Hilbert-Schmidt operator $H_a$ defined by
\be
\label{hs}
(H_a(x))_n:=a_n x_n \ , \ \ x\in\ell^2 \, .
\ee
Then the Bochner-Minlos Theorem
leads to the conclusion that a typical sequence $x$ in the support of
the measure must
satisfy
\be
\label{minlos}
\sum^{\infty}_{n=1}a_n^2 x_n^2<\infty\ ,
\ee
which is certainly not true for a sequence behaving
asymptotically like $\sqrt{2\rho\ln n}$, if we choose
appropriately $a\in \ell^2$. 
However,
the Bochner-Minlos theorem does not forbid the appearance of
a subsequence behaving
asymptotically even worse than $\sqrt{2\rho\ln n}$. 
Therefore proposition~\ref{prop1} means that in  
a typical sequence $\{x_n\}$ no subsequence $\{x_{n_k}\}$ can be found 
such that, $\forall n_k$, 
$|x_{n_k}|> \sqrt{2(1\!+\!\epsilon)\rho\ln n_k}$,      
for any arbitrary but fixed $\epsilon$ greater than zero, 
and that one is certainly found if
$\epsilon$ is taken to be zero.
But this subsequence is rather sparse,
as demanded by Bochner-Minlos Theorem; from a stochastic
point of view the occurrence of values $|x_n|$  greater 
than $\sqrt{2\rho\ln n}$
is a  rare event. The typical sequence in the support is one that is 
generated with a probability distribution given by the measure. The measure 
in this case is just a product of identical Gaussian measures in $\R$,
so the typical sequence is one obtained by throwing a ``Gaussian dice''
an infinite number of times.

Notice  that a typical $\mu_{\rho'}$ sequence can be
obtained from a
\ms\ typical sequence simply by multiplying by $\sqrt{\rho'/\rho}$.
This follows from the fact that the map
\mbox{$x\mapsto \sqrt{\rho'/\rho}\ x$} is a
isomorphism of measure spaces $(\R^{\Ni}, \ms) \rightarrow 
(\R^{\Ni},\mu_{\rho'})$.


\subsection{The White Noise Measure}
\label{s2.4}

We consider now the so called ``white noise''
measure, which in some sense is the continuous analogue 
of the previous case\cite{HPS}.
Again, we will look for convenient sets of measure one, in the 
sense given in section \ref{s2.3}. This will be achieved by a 
proper choice of
random variables, i.e. measurable functions, which will
reduce the present case to the previous discrete one.

As mentioned in section \ref{s2.1} the d+1-dimensional white noise  is 
the Gaussian measure $\ns$
with Fourier transform $\chi_{\s}(f) = \exp(-\s/2(f,f))$.
Notice that here $\s $ has dimensions of inverse mass squared.

The Euclidean group $\cal E$ acts on \sd, \sdd\ and (unitarily) on
\break $L^2(\sdd,d\ns)$ respectively by
\begin{eqnarray}
\label{c3}
\left( \tilde \varphi_g f\right)(x) & = & f(g^{-1} x)\, ,\nonumber \\
\left( \varphi_g\f\right)(f) & = & \f (\tilde \varphi_{{g^{-1}}} f)\, ,\\
\left( U_g \psi\right)(\f) & = &\psi 
\left( \varphi_{{g}}\f\right)\, ,  \nonumber
\end{eqnarray}
where $g\in \cal E$, $g\,x$ denotes the standard
action of $\cal E$ on $\R^{d+1}$ by translations,
rotations and reflections, $f\in\sd$, $\f\in\sdd$ 
and $\psi\in L^2(\sdd,d\ns)$.

It is  easy to see that a subgroup of translations in a fixed
direction, say the time direction,
is mixing. One just has to consider the set with dense span of
$L^2$-functions of the form $\exp (i\,\f(f))$, $f\in\sd$ and use the
Riemann-Lebesgue Lemma to prove that
\be
\label{c4}
\lim_{t \to \infty}\int f(x_0+t,\ldots,x_{d})\,
 g(x_0,\ldots,x_{d})\, d^{d+1} x = 0\ ,
\ \forall f,g\in \sd\, .
\ee
This implies that the measures $\ns$ and $\mu_{\sigma '}$
for $\s \neq \s'$ have disjoint supports,
even though the Bochner-Minlos theorem gives us for the
support in both cases 
an extension
of $L^2(\R^{d+1},d^{d+1}x)$ through an Hilbert-Schmidt operator 
(see section \ref{s2.3}). 
Since the choice of
a  complete $( \ , \ )$-orthonormal system
$\{f_n\}$ gives us a isomorphism
of measure spaces
\ba
\label{245}
(\sdd, \ns) &\rightarrow& 
(\R^{\Ni}, \ms)_{|_{\rho = \s}} \nonumber \\
\phi &\mapsto& \{\phi(f_n)\}\ ,  
\ea
for every such basis one can find sets of the type
of those found in the previous subsection
and which put in evidence the
mutual singularity of $\ns$ and
$\mu_{\sigma'}$. However, for the 
convenience of our analysis of
free massive scalar fields in the 
next section, let us study the 
$x$-behavior of typical white noise
configurations $\phi$. Since
\be
\label{246}
\delta_x \ : \ \phi \mapsto \phi(x)
\ee
is not a good random variable, we fix in ${\R}^{d+1}$ a 
family of non-intersecting cubic boxes, $\{B_j\}_{j = 1}^\infty$,
with sides of length $L$. Then the mean value of $\phi$ over
$B_j$ is a well defined random variable
\be
\label{247}
F_{B_j} \ : \ \phi \mapsto F_{B_j}(\phi)\equiv
\phi(f_j) = {1 \over L^{d+1}} \int_{B_j} \phi(x) d^{d+1}x
\ ,
\ee
where $f_j$ denotes the characteristic function
of the set $B_j$ divided by the volume $L^{d+1}$, and the
map 
\ba
\label{248}
\sdd &\rightarrow& \R^{\Ni}  \nonumber \\
\phi &\mapsto& \{\phi(f_j)\}  
\ea
defines (by push-forward) a measure on $\R^{\Ni}$ 
of the form (\ref{3.5}) with
$\rho = \s / L^{d+1}$. This can be seen from the fact that
\be
\label{249}
\int_{\sdd} d\ns \exp \Bigl({i 
\sum_{j=1}^\infty y_j \phi(f_j)}\Bigr) = 
\exp \Bigl(-{ {\s \over L^{d+1}}
{ \sum_{j=1}^\infty y^2_j  \over 2}}\Bigr) \ .
\ee
We then conclude from section \ref{s2.3} that the sets
\ba
\label{250}
W_{\s}^{\epsilon}
:=\bigl\{\f\! & \in & \!\sdd\,:\  \exists 
N_{\f}\in \N\ \ \ \mbox{s.t.}\ \nonumber \\ 
 |\f(f_n)|\!& < & \!\sqrt{2(1+\epsilon)(\s/L^{d+1})\ln n},
\ \ \mbox{for}\  n\geq N_{\f} \bigr \}  
\ea
have $\ns$-measure one for every positive $\epsilon$, and that for 
$\s'\! >\! \s$ an $\epsilon(\s')\!>\!0$ can be found such that
$\mu_{\sigma'}(W_{\s}^{\epsilon(\s')})=0$. This
shows that the supremum of the mean value of $\phi$ over $N$
boxes with volume $L^{d+1}$ goes like $\sqrt{2(1\!+\!\epsilon)
(\s/L^{d+1})\ln N}$.
A white noise with a bigger variance $\s' > \s$ has the
latter behavior on larger boxes with volume 
$L'^{d+1} = {\s' \over \s} L^{d+1}$.


\section{Quantum Scalar Field Theories}
\label{s3}


\subsection{Constructive Quantum Scalar Field Theories}
\label{s3.1}

Here we recall briefly 
some aspects of constructive quantum
field theory that will be relevant for the
next subsection\cite{glijaf}.  A quantum scalar field theory
on $d+1$-dimensional (flat Euclideanized) space-time
is a measure $\mu$ on
$\sdd$ with Fourier transform $\chi$  (generating
functional or $\chi(f) = Z(-if)/Z(0)$ in theoretical
physics terminology) satisfying the
Osterwalder-Schrader (OS) axioms. We will be
interested in the axioms which state the Euclidean
invariance of the measure (OS2) and ergodicity
of the action of the time translation subgroup  (OS4), i.e.
for $g=T_t : \ T_t(t',{\bf x}) = (t' + t, {\bf x})$, 
\be
\label{ergo2}
\lim_{\tau \rightarrow \infty} {1 \over 2 \tau} \int_{-\tau}^\tau
dt \psi(\vff_{T_t} \phi_0) =_{a.e.} \int_{\sdd} \psi(\phi) d\mu(\phi)  \ .
\ee
The action of the Euclidean group on $\ss$,
$\sdd$ and $L^2(\sdd,d\mu)$ is defined as in section \ref{s2.4}.
So OS2 states that $\vff_{g*} \mu = \mu$ for all
elements $g$ of the Euclidean group $\cal E$. Notice that
OS2+OS4 imply  that     ergodicity under the
subgroup of time translations is equivalent to
ergodicity under the full Euclidean group.
The vacua of the theory correspond to
Euclidean invariant vectors on  $ L^2(\sdd,d\mu)$ and the
axioms OS2 and OS4 imply that the vacuum is
unique and given by the constant function.
Examples of measures satisfying the OS axioms are
the Gaussian measures $\nm$ corresponding to
free massive quantum scalar field
theories (see eqs. (\ref{2.3c},\ref{2.3d})).


\subsection{Support of Free Scalar Field Measures}
\label{s3.2}

Since the Euclidean group $\cal E$ acts ergodically
on $(\sdd, \mu_{C_m})$ we conclude that the measures
$ \mu_{C_m}$ and $ \mu_{C_{m'}}$ (see eq. (\ref{2.3d})) with $m \neq m'$
must have disjoint supports. Like in section
\ref{s2.4} we will characterize the difference
of supports in terms of the mean value of $\phi$
over a region with volume $L^{d+1}$. Before going
into the details of the calculations notice that
the map
\ba
\label{321}
(\sdd, \ns) & \rightarrow & (\sdd, \mu_{C_m}) \nonumber \\
\phi & \mapsto & \left[\s (-\Delta + m^2) \right]^{-1/2} \phi
\ea
is an isomorphism of measure spaces which maps typical
white noise configurations to typical $\nm$-configurations.
Heuristically this means that for big distances
($\Delta x >> {1 \over m}) $ or small momenta the
correlation imposed by the kinetic term in the
action is lost and the typical configurations approach
those of white noise with $\s = {1 \over m^2}$.
Let us now obtain a formal derivation of this
fact as far as the $x$-space behavior of
typical configurations of free massive scalar fields
is concerned.

Consider the same random variables $F_{B_j}$ as
in section \ref{s2.4}  but, in order to
eliminate the correlation, the cubic boxes $B_j$  will be
chosen centered in the points $x^j =(x_0^j, ... , x^j_d) =
({j^2 \over m} , 0, ... , 0)$
and with sides parallel to the
coordinate axes. The push-forward of $\mu_{C_m}$ with respect
to the map
\ba
\label{push}
\sdd &\rightarrow& \R^{\Ni}  \nonumber \\
\phi &\mapsto& \{\phi(f_j)\}  
\ea
is a Gaussian measure $\mu_{M_m}$ in $\R^{\Ni}$ with covariance matrix 
${M_m}$ given by
\ba
\label{Mm}
(M_m)_{jl}&=&C_m(f_j,f_l) =\\ 
&=&\left({2 \over \pi}\right)^{d+1}{1 \over L^{2(d+1)}}
\int_{\R^{d+1}} d^{d+1} p\,  e^{i{p_0 \over m}(j^2 - l^2)}{1 \over p^2 + m^2} 
\prod_{k=0}^d{{\sin}^2(p_kL/2) \over p_k^2}\, .\nonumber
\ea
Let us denote the (constant) value of the diagonal elements of
$M_m$ by $C_m^{L}$, i.e.
\ba
\label{CLm}
C_m^{L} &:=& (M_m)_{ii} =\\  
&=&\left({2 \over \pi}\right)^{d+1}{1 \over L^{2(d+1)}}
\int_{\R^{d+1}} d^{d+1} p \, {1 \over p^2 + m^2} 
\prod_{k=0}^d{{\sin}^2(p_kL/2) \over p_k^2} \ .\nonumber
\ea
\begin{prop}
\label{prop2}
The set
\ba
\label{Yme}
Y^{\epsilon}_{(m)}&:=&\{ \phi \in \sdd  : \exists N_{\phi}\in\N 
\ \ {\rm s.t.}\nonumber \\  
&|\phi(f_n)|& < 
\sqrt{2 (1+\epsilon)C_m^{L} \ln n} \ , \ n \geq N_{\phi} \} 
\ea
has $\nm$-measure one for any $\epsilon > 0$.
\end{prop}
Like in Section~III.2, we will show that the $\mu_{M_m}$-measure
of the image of $Y^{\epsilon}_{(m)}$ in $\R^{\Ni}$ is one. To prove
this we will relate the measure $\mu_{M_m}$ with a diagonal 
Gaussian measure of the form (\ref{3.5}).
Let $\mu_{C^L_m}$ be the Gaussian measure in $\R^{\Ni}$ with diagonal 
covariance matrix $C^L_m \delta_{ij}$.
\begin{lem}
\label{mac}
The measures $\mu_{M_m}$ and $\mu_{C^L_m}$ are mutually absolutely 
continuous, i.e. have the same zero measure sets.
\end{lem}
To prove the lemma we will rely on Theorem~I.23 in pag.~41 of Ref.~1
(see also Theorem~10.1 in pag.~160 of Ref.~32),
which gives necessary
and sufficient conditions for two covariances to give rise to mutually 
absolutely continuous Gaussian measures. In our case, since 
the covariance of $\mu_{C^L_m}$ is proportional to the identity, it is
sufficient to show that ({\it i}\/) the operator 
$T:=M_m-C^L_m{\rm \bf 1}$ is Hilbert-Schmidt 
and ({\it ii}\/) the operator $M_m$ is bounded, positive 
with bounded inverse in $\ell^2$. 
Let us first prove that $T$ is Hilbert-Schmidt. The matrix
elements of $T$ are $T_{ii}=0$
and $T_{jl}=(M_m)_{jl}$, for $j\not = l$. One can 
see from (\ref{Mm}) that the off-diagonal elements of $(M_m)_{jl}$
are the values at the points $j^2-l^2$ of the Fourier transform of a real
function $f$. Explicitly,
\be
\label{Mmf}
(M_m)_{jl}=\int_{\R} d\nu_0\, e^{i\nu_0(j^2-l^2)} f(\nu_0)\ ,
\ee
where 
\ba
\label{f}
f(\nu_0)&:=&
\left({2 \over \pi}\right)^{d+1}{1 \over m^{d+3} L^{2(d+1)}}
{{\sin}^2(mL\nu_0/2) \over \nu_0^2}\ \cdot\nonumber \\
&\cdot&\int_{\R^{d}} d^d\nu\, {1 \over 1+\nu_0^2+\sum_1^d \nu_k^2} 
\prod_{1}^d{{\sin}^2(mL\nu_k/2) \over \nu_k^2}\ .
\ea
Since both $f$ and its derivative $f^{\prime}$ are $L^1$, one gets for the
Fourier transform $\tilde f$ of $f$:
\be
\label{ftilde}
| \tilde f(t)|={|\widetilde{f^{\prime}}(t)| \over |t|}\ ,
\ee
with $\widetilde{f^{\prime}}$ continuous, bounded and approaching zero 
at infinity. Therefore exists $A>0$ such that
\be
\label{boundM}
|(M_m)_{jl}|^2 \leq {A \over (j^2-l^2)^2}\ , \ \ {\rm for}\ j\not = l\ ,
\ee
and therefore
\be
\label{sumconv}
\sum_{j,\, l} |T_{jl}|^2 \leq A \sum_{j\not = l}
{1 \over (j^2-l^2)^2} < \infty\ ,
\ee
thus proving that $T$ is Hilbert-Schmidt.
Let us now prove ({\it ii}\/). The operator $M_m$ is a positive operator
on $\ell^2$ since it is given by the restriction of the positive
covariance $C_m$ on $L^2(\R^{d+1})$ to the linearly independent
system $\{f_j\}_{j\in\Ni}$. Positivity of $M_m$ and the fact that
$M_m=C^L_m{\bf 1}+T$, $T$ being compact (in fact Hilbert-Schmidt), 
implies that $M_m$ is bounded, has a trivial kernel and therefore
is invertible with bounded inverse (see e.g. Theorem 4.25 and open
mapping theorem in Ref.~34).
Proposition~2 now follows, given the characterization
of the support of $\mu_{C^L_m}$ one gets from section~\ref{s2.3}.~$\Box$
 
Using the fact that
\be
\label{delta}
{1\over \alpha\pi} {\sin^2(\alpha p)\over p^2}
\ee
tends to $\delta(p)$ when $\alpha\to\infty$, one sees from (\ref{CLm}) that
\be
\label{CLmto}
\lim_{L\to\infty} L^{d+1}C_m^L={1\over m^2}\ .
\ee
Thus, comparing (\ref{Yme}) with (\ref{250}) we see that,
in accordance with the discussion above, when averaged over widely
separated large boxes ($L >> {1 / m}$), the typical
free field distribution 
approaches white noise with $\s = {1 / m^2}$.

The explicitness of the mutual singularity of $\mu_{C_m}$ and 
$\mu_{C_{m^{\prime}}}$, with $m\not = m^{\prime}$ now follows
easily from (\ref{difset}), (\ref{Yme}) and the fact that
$C_m^{L}$ is a monotonous (decreasing) function
of $m$.


\section{Properties of the Ashtekar-Lewandowski Measure
on Spaces of Connections}
\label{s4}


\subsection{Ergodic Action of the Group of Diffeomorphisms}
\label{s4.1}

The diffeomorphism-invariant Ashtekar-Lewandowski measure
$\mu_{AL}$ on the space of connections modulo gauge
transformations  $\ag$ over the manifold $\Sigma$
plays an important role in rigorous attempts to
find a  quantization of canonical
general relativity. In the present section we
study properties of this measure. We show that
$Dif\!f\!{}_{{}_0}(\Sigma)$,
the connected component of the group of diffeomorphisms
of (the connected analytic manifold) $\Sigma$, acts ergodically on $\agb$. 
In the next subsection we
also obtain results concerning the properties
of the support of $\mu_{AL}$.

Let us recall the definition of
$\mal$. We denote by a hoop $[\alpha]$
in $\S$ the equivalence class of
(piecewise analytic) loops $\{\widetilde \a \}$
based on $x_0 \in \S$ such that
\be
\label{4.1}
U_{\wt \a}(A) = U_\a(A)   \ ,   \forall A \in {\cal A}\ ,
\ee
where $ U_\a(A)$ denotes the holonomy corresponding
to the loop $\a$, the connection $A$ and a chosen
point in the fiber over $x_0$ of the (fixed) principal $G$-bundle
over $\S$ and $\cal A$ is the space of all   
connections on this bundle. The set of all hoops forms a group
$\cal HG$, called the hoop group\cite{ashlew}. We note that
for all $G=SU(N)$ with $N\geq 2$ the group $\cal HG$
does not depend on $N$ nor on the principal
bundle\cite{ashlew}.
Throughout the present section we will assume $G$ to
be a compact connected Lie group.
 It is well known that
a connection $A$ defines through $U_{.}(A)$ a homomorphism
from  $\cal HG$ to the gauge group $G$: $[\alpha] \ \mapsto \
U_\alpha(A)$. In fact, the space
$\ag$ is in a natural bijection with the 
space of all appropriately smooth homomorphisms
of this type, modulo conjugation at the base point\cite{pie,lew2,cae}. 
On the other hand, and as expected
from the example of scalar fields, the measure
$\mal$ lives in a space bigger than the classical
space $\ag$. This is the space of ${\it all}$,
not necessarily smooth or even continuous, 
homomorphisms from $\hg$ to $G$ modulo conjugation, denoted by
$\agb$~\cite{ashish,ashlew}. The space $\ag$ of smooth classes
$[A]$ was shown to be of zero measure in $\agb$~\cite{marmou}.
In the next subsection we will deepen this result.

The  measure
$\mal$ is, as in the scalar field case, completely specified by giving the
result of integrating the so-called cylindrical
functions, which in this case are 
gauge-invariant functions of a finite number
of parallel transports along (analytic embedded) edges
\be
\label{4.2}
f(\Ab)=F(\Ab(e_1), ... ,\Ab(e_n))   \ ,
\ee
where different edges may intersect only on the ends
and $\Ab\in
\bar {\cal A}$, the space of all connections realized as parallel transports
in a natural way\cite{ashlew2,almmt}.
The measure $\mal$ is then defined by
\be
\label{4.3}
\int_\agb d\mal f(\Ab) = \int_{G^n} dg_1...dg_n 
F(g_1, ... ,g_n)   \ ,
\ee
where $dg$ is the normalized Haar measure on $G$.

The group $Dif\!f\!{}_{{}_0}(\Sigma)$ has a natural action on $\bar {\cal A}$ 
which leaves $\mal$ invariant
\be
\label{4.4}
\vff^{\ast}\Ab(e):=\Ab(\vff\!\cdot\! e).
\ee
As we have seen in section \ref{s2.2} this action 
induces a unitary action of 
$Dif\!f\!{}_{{}_0}(\Sigma)$ on \mbox{$L^2(\agb,d\mal)$}
\be
\label{4.5}
\left( U_{\vff}f\right)=f(\vff^{\ast}\Ab).
\ee
Consider now the so called 
{\em spin-network states}\/\cite{R-S,baez1,baez2} $\{T_s\}$
, indexed by triples
$s=(\gamma,\pi,c)$, where $\gamma$ is a graph, 
$\pi\!:=\!(\pi_1,\ldots,\pi_n)$ is a labeling of the edges of $\gamma$
with nontrivial irreducible representations $\pi_i$ of $G$ and 
$c\!:=\!(c_1,\ldots,c_m)$ is a labeling of the vertices
$v_1,\ldots,v_m$ of $\gamma$ with {\em contractors} $c_j$, i.e. nonzero
intertwining operators from the tensor product of the representations
corresponding to the incoming edges at $v_j$ to the tensor product of
the representations associated with the outgoing edges.
The unitary action of $Dif\!f\!{}_{{}_0}(\Sigma)$ on the spin-network
states is particularly simple, being given by
\be
U_{\vff}T_{\gamma,\pi,c}=T_{\vff\gamma,\vff\pi,\vff c}
\ee
or, in short
\be
U_{\vff}T_s=T_{\vff s}\, ,
\ee
where $\vf\gamma$ is the image of the graph $\gamma$ under the
diffeomorphism $\vf$ and $\vf\pi$ and $\vf c$ are the 
corresponding representations and contractors associated with 
$\vf\gamma$. A crucial property is that the contractors can be chosen
in such a way that the spin-network states 
form an orthonormal basis on 
\mbox{$L^2(\agb,d\mal)$}\cite{baez1,baez2,tho-loop}.
We will assume that this has been done to 
prove the following 
\begin{theo}
The group $Dif\!f\!{}_{{}_0}(\Sigma)$ acts ergodically on 
$\agb$, with respect to the  measure $\mal$.
\end{theo}
To prove the theorem, we will show that 
the only $Dif\!f\!{}_{{}_0}(\Sigma)$-invariant vector on 
\mbox{$L^2(\agb,d\mal)$} is the constant function. 
Therefore there can be no measurable 
$Dif\!f\!{}_{{}_0}(\Sigma)$-invariant subsets of $\agb$ with measure
different from zero or one.

Using the completeness of the spin-network states,
every $\psi\in L^2(\agb,d\mal)$ 
can be represented in the form
\be
\label{4.6}
\psi=\sum_s c_s T_s\, ,
\ee
where no more than countably many coefficients $c_s$ are nonzero. 
Since for any $s$ and any diffeomorphism $\vf$, $T_{\vff s}$ 
belongs to the
same orthonormal basis ($\vf s\neq s 
\Rightarrow T_{\vff s}\perp T_s$),
we conclude that $\psi$ in (\ref{4.6}) is
$Dif\!f\!{}_{{}_0}(\Sigma)$-invariant if and only if
\be
c_{\vff s}=c_s \ \ \ \forall \vf\in Dif\!f\!{}_{{}_0}(\Sigma).
\ee
A $L^2$-vector cannot have more than finitely many equal coefficients,
and since for every nontrivial spin-network 
(with nontrivial graph
and representations) there is an infinite 
(actually uncountable) number of (distinct) spin-networks in the orbit
\be
Dif\!f\!{}_{{}_0}(\Sigma)\, s:=\{\vf s\, , \ \vf\in 
Dif\!f\!{}_{{}_0}(\Sigma)\}\, ,
\ee
we conclude that an invariant $\psi$ in (\ref{4.6}) is necessarily 
constant almost everywhere.~$\Box$

$ $ From the proof it follows that if $H$ is a subgroup of 
$Dif\!f\!{}_{{}_0}(\Sigma)$
s.t. the $H$-orbit $Hs$ through every nontrivial spin-network $s$
is infinite, then $H$ acts ergodically on $\agb$. So for example
\begin{cor}
If $\ \Sigma=\R^N$ or $\ \Sigma=T^N$ the measure space   
$(\agb,\mal)$ admits the ergodic action of subgroups $H$ 
of $Dif\!f\!{}_{{}_0}(\Sigma)$ isomorphic to $\Z$.
\end{cor}
For $\Sigma=\R^N$ we take the $\Z$-subgroup of $Dif\!f\!{}_{{}_0}(\Sigma)$
generated by $\vf_0\, $:
\be
\label{4.7}
\vf_0(x^1,\ldots,x^N)=(x^1+\omega^1,\dots,x^N+\omega^N)\, ,
\ee
where $(\omega^1,\dots,\omega^N)$ is a fixed non-vanishing 
vector in ${\R}^N$ (recall that spin-networks are defined here 
only for finite graphs).
If $\Sigma=T^N$ we take in (\ref{4.7}) the vector  
$(\omega^1,\dots,\omega^N)$ to have (at least) two irrational
and incommensurable components, where $(x^1,\ldots,x^N)$
are now mod~1 coordinates of $T^N$. In both cases for 
every nontrivial spin network $s$, $\{\vf_0^n s,\ n\in \Z\}$
contains an infinite number of distinct spin networks and so 
the group $\{\vf_0^n,\ n\in \Z\}$ acts ergodically on 
$(\agb,\mal)$.~$\Box$


\subsection{Support Properties}
\label{s4.2}

As we mentioned in the beginning of section \ref{s4.1} the
space $\ag$ of smooth connections modulo gauge transformations
is contained in a zero measure subset of $\agb$, 
the space where the
measure $\mal$ is naturally defined\cite{ashish,ashlew,marmou}. 
The latter space is naturally identified
with the space of all (not necessarily continuous) homomorphisms from
the hoop group $\cal HG$ to the gauge group $G$
modulo conjugation at the base point\cite{ashlew2}.
In the present subsection we deepen the result
of Ref.~8
by showing that the parallel transport
of a $\mal$-typical connection $\Ab \in \overline {\cal A} $
along an edge $e$ leads to a nowhere continuous map
\be
\label{52.1}
g(\cdot) \ : \ [0,1] \ \rightarrow \ G \ .
\ee
Indeed let $e \ : \ [0,1] \ \rightarrow \ \S$ be
an arbitrary edge and consider for $s \in [0,1]$
the part of the edge $e_s$ given by
$$
e_s(t) = e(st) \ , \ t \in [0,1] \ .
$$
We then have a map 
\ba
\label{52.2}
v \ : \ {\overline {\cal A}} \ &\rightarrow& G^{[0,1]} \\
                     \Ab \ &\mapsto& \ v_{\Ab} \ ,
\qquad  v_{\Ab}(s) := \Ab(e_s) \ ,     \nonumber
\ea
where $G^{[0,1]}$ denotes the space of all maps from $[0,1] $
to $G$.
By choosing in $G^{[0,1]}$ the standard product space topology
and as algebra of measurable sets the Borel $\sigma$-algebra
the map $v$ becomes measurable.

It is easy to see that, due to the properties of the
Haar measure, the push-forward $v_*\mal$ of $\mal$\cite{foot2}
to $G^{[0,1]}$ is a product of Haar measures, one for each point 
$s \in [0,1]\, $:
\be
\label{52.3}
d\nu(g(\cdot)) = v_* d\mu_{AL} = \prod_{s\in[0,1]}dg(s) \ .
\ee
The main result of this subsection is the following

\begin{theo}
The measure $\mal$ is supported on the set $W$ of 
all connections $\Ab$ such that $v_{\Ab}$ is 
everywhere discontinuous as a map from $[0,1]$ to $G$.
\end{theo}
Since $W = v^{-1}(W_1)$, where
\be
\label{52.4}
W_1 = \{ g(\cdot) \in G^{[0,1]} \ , \ \ \mbox{s.t.}\ \
g(\cdot) \ \hbox{\rm is nowhere continuous} \} \ ,
\ee
it is sufficient to prove that the complement of $W_1$,
\be
\label{52.5}
W_1^c = \{ g(\cdot) \in G^{[0,1]} \ \ : \ \exists s_0 \in [0,1]\
 \ \mbox{s.t.}\ \  g(\cdot) \ \hbox{\rm is continuous at $s_0$} \} \ ,
\ee
is contained in a zero $\nu$-measure subset of $ G^{[0,1]}$.
Consider the sets
\be
\label{52.6}
\Theta_U = \{g(\cdot) \in G^{[0,1]} \ : \ \exists I \ \ \mbox{s.t.}\  \ g(I)
 \subset U\} \ , 
\ee
where $U$ is a (measurable) subset of $G$ with $0<\mu_H(U)<1$ ($\mu_H(U)$ 
denoting the 
Haar measure of $U$) and $I$ is an open subset of $[0,1]$.
We need the following

\begin{lem}
For every $U\subset G$ with  $0<\mu_H(U)<1$ the set $\Theta_U$
is contained in a zero measure subset of $G^{[0,1]}$.
\end{lem}
To prove the lemma recall that the open balls
\be
B(q,1/m)=\{s\in[0,1]\ : \ |s-q|<1/m\}
\ee
with rational $q$ and integer $m$ are a countable basis for the topology
of $[0,1]$. Thus $\Theta_U$ is the countable union of the sets
\be
\Theta_{U,q,m} := \{g(\cdot) \in G^{[0,1]} \ : \   g(B(q,1/m))
\subset U\}, \ \ q\in\Q, \ m\in\N\, .
\ee 
It is easy to construct zero measure subsets containing 
$\Theta_{U,q,m}$.
For this fix an infinite sequence $\{s_i\}_{i=1}^\infty \subset B(q,1/m)$ 
of distinct points, $s_i \neq s_j$ for $i \neq j$. Then the set
$Z = \{g(\cdot) \in G^{[0,1]} \ : \ g(s_i) \in U \ , \  i \in \N\}$ contains 
$\Theta_{U,q,m}$ and has zero measure:
$$
\nu(Z) = \lim_{n \rightarrow \infty}(\mu_H(U))^n = 0 \ .
$$
$ $From the $\sigma$-additivity of $\nu$
we conclude that, for every subset $U \subset G$ with
Haar measure less than one, the set $\Theta_U$ is contained
in a zero $\nu$-measure subset.~$\Box$

We can now conclude the proof of the theorem.
Let us choose a number $r$,  \mbox{$0\! <\! r\! <\! 1$} and
a finite open covering $\{U_i\}_{i=1}^k$ 
of the compact group $G$ with $\mu_H(U_i) = r  \ , \ i=1, ... , k$.
This is clearly possible since we can take a neighborhood
of each point with measure $r$ and then take a finite sub-covering.
Consider $g(\cdot) \in W_1^c$. Then there exists a $s_0 \in [0,1]$ such that
$g(\cdot)$ is continuous at  $s_0$. Let $i_0$ be such that $g(s_0) \in
U_{i_0}$. Continuity implies that there exists a neighborhood
$I$ of $s_0$ such that $g(I) \subset U_{i_0}$ and therefore
we have
$ W_1^c \subset \cup_{i=1}^k \Theta_{U_i}\, .\ \Box $


\section{Conclusions and Discussion}
\label{s6}

The knowledge of the support of measures on
infinite dimensional spaces, used in quantum field theory,
gives a grasp on the behavior of typical quantum 
field configurations associated with these measures. This may be important
for a better understanding, both from the physical
and mathematical points of view, of problems afflicting 
interacting theories like the problem of divergences.

The Bochner-Minlos theorem is very effective in capturing linear 
properties of the 
support of measures on the space ${\cal S}'(\R^{d+1})$ of 
quantum scalar field configurations (the support is a linear subspace
of  ${\cal S}'(\R^{d+1})$ spanned by configurations with
a given norm finite). It allows e.g. to distinguish
the support of the measure $\mu_{C_m}$, corresponding
to  free scalar field theory with mass $m$, from 
that of the white noise measure.
However this theorem would predict the same support
for the measures $\mu_{C_m}$ and $\mu_{C_{m'}}$ with 
$m \neq m'$ even though  these measures must have disjoint
supports. The latter is due to the fact  that any one parameter subgroup of 
translations of $\R^{d+1}$ acts ergodically on 
 ${\cal S}'(\R^{d+1})$
with respect to
both measures. The above two claims do not contradict and rather
complement each other since the subset of a ``support'' which
is thick, i.e. such that any measurable subset of its complement has 
measure zero,
is also a (finer) support for the same measure. 
So, to distinguish between 
$\mu_{C_m}$ and $\mu_{C_{m'}}$, one has to find properties
of the supports which complement those given by
the Bochner-Minlos theorem.
Support properties
of  $\mu_{C_m}$ were studied in Refs.~26 and 27 
(see also Refs.~28, 29 and 30). 
Our goal in section \ref{s3.2} consisted in 
showing that a simple system of random variables
can be used to study the difference of supports of
$\mu_{C_m}$ and $\mu_{C_{m'}}$ for large distances.
This simplifies part of the results of Refs.~26 and 27
and makes them physically more transparent.

The use of the AL measure in attempts to construct
a quantum theory of 
gravity\cite{ashish}${}^{\hbox{-}}$\cite{tho4}$
{}^{\rm ,}$\cite{RS}${}^{\hbox{-}}$\cite{lew3} 
justifies the importance to
study its properties. In section \ref{s4} we improved
the results of Ref.~8.
We fixed an arbitrary edge
$e$, or equivalently a piecewise analytic curve on $\Sigma$,
and considered the random variables given by $\Ab \mapsto \Ab(e_s)$,
where $\Ab$ is a quantum connection, $e_s$ represents the
edge $e$ up to the value $s \in [0,1]$ of the 
parameter and $\Ab(e_s)$ is the parallel transport
corresponding to $\Ab $ and $e_s$.
These random variables were suggested to the authors by Abhay
Ashtekar and were motivated by studies of the
volume operator in quantum gravity\cite{RS}${}^{\hbox{-}}$\cite{vol}.
Varying $s$ we obtain a (measurable) map $v$ 
from the space of quantum connections $\overline {\cal A}$
to the space $G^{[0,1]}$ of maps, $g(\cdot) \ : \ [0,1] \ \rightarrow
\ G$. The AL measure 
becomes, by push-forward, an infinite product of Haar measures. 
Using this we have shown that for a typical $\mal$-connection
$\Ab$ the parallel transport $\Ab(e_s)$ is everywhere discontinuous.

We also showed that $Dif\!f\!{}_{{}_0}(\Sigma)$ acts
ergodically on $\agb$ with respect to the 
AL measure. The importance of this stems from the
fact that in quantum gravity
one has to solve the diffeomorphism constraint and therefore
naively one would have to take functions on the quotient
$$
\agb / Dif\!f\!{}_{{}_0}(\Sigma)   \ .
$$
The ergodicity of the action of $Dif\!f\!{}_{{}_0}(\Sigma)$
on $\agb$ implies that the only solution
to the diffeomorphism constraint in $L^2(\agb, d\mu_{AL})$
is the constant function. This explains why in Ref.~11, 
and even though $\agb$ is compact,
one had to use distributional elements to
solve the diffeomorphism constraint. 
If the action were not ergodic one would have diffeomorphism 
invariant measurable subsets of $\agb$ (pre-images of sets in
$\agb / Dif\!f\!{}_{{}_0}(\Sigma)$) with 
$\mal$-measure different from zero
or one. The characteristic functions of these sets would provide
$L^2$ solutions to the diffeomorphism constraint.


\nonumsection{Acknowledgments}
\noindent
We would like to thank Abhay Ashtekar,
Donald Marolf and Jerzy Lewandowski for useful discussions.
JMM and TT would like to thank ESI, where a significative
part of their contribution to this work was
accomplished. JMM and JMV were in part supported
by PCERN/P/FAE/1111/96, CENTRA/IST and
CENTRA/Algarve. TT was supported in part by 
DOE-Grant DE-FG02-94ER25228 of Harvard University.
JMV was supported by grant PRAXIS XXI BD/3429/94.


\newpage


\nonumsection{References}
\bigskip

\noindent\hangafter=1\hangindent=7pt
${}^{\rm a)}$Present address: Departamento de F\'{\i}sica, Instituto
Superior T\'ecnico,
Av. Rovisco Pais, 1096 Lisboa, Portugal.
e-mail: jmourao@galaxia.ist.utl.pt
\bigskip

\noindent\hangafter=1\hangindent=7pt
${}^{\rm b)}$Present address: Max-Planck-Institut f\"ur Gravitationphysik,
Albert-Einstein-Institut,
Schlaatzweg 1, 14473 Potsdam, Germany.
e-mail: thiemann@aei-potsdam.mpg.de
\bigskip

\noindent\hangafter=1\hangindent=7pt
${}^{\rm c)}$Present address: Unidade de Ci\^encias Exactas e Humanas,
Universidade do Algarve,
Campus de Gambelas, 8000 Faro, Portugal.
e-mail: jvelhi@ualg.pt



\begin{thebibliography}{999}
\bibitem{foot1} The support of the
measure $\mu$ on $M$ is any (not necessarily
measurable) subset $X \subset M$ such that
any measurable subset $Y$ of the
complement $Y \subset X^c$ has measure zero.
Then $\mu$ can be identified with the $\sigma$-additive
measure $\mu'$ on $X$ defined on sets of the
form $X \cap Y$, with $Y$ measurable on $M$, by
$\mu'(X \cap Y) := \mu(Y)$.
\bibitem{foot2} Here
$\mal$ is considered on $\overline {\cal A}$. It is  defined
analogously to eq. (\ref{4.3}) with the difference that
the  functions of
parallel transports are not necessarily gauge invariant (see Ref.~7).
The push-forward of this measure to $\agb = {\overline {\cal A}}/
{\overline {\cal G}}$ gives the Ashtekar-Lewandowski measure
on this space.
\end{thebibliography}
\end{document}